\documentstyle[12pt]{article}
\begin{document}
\title{Cosmological model with movement in fifth dimension}
\author{ W. B. Belayev \thanks {dscal@ctinet.ru}\\
\normalsize Center for Relativity and Astrophysics,\\ \normalsize
185 Box , 194358, Sanct-Petersburg, Russia }

\maketitle

\begin{abstract}

  Presented cosmological model is 3D brane world sheet moved in extra
dimension with variable scale factor. Analysis of the geodesic
motion of the test particle gives settle explanation of the
Pioneer effect. It is found that for considered metric the
solution of the semi-classical Einstein equations with various
parameters conforms to isotropic expanded and anisotropic stationary
universe.

\end{abstract}

\section{Introduction}

  Although conception of expanded universe is predominant, stationary
model is also suggested \cite{1}, \cite{2}. Godel \cite{3}
investigated a zero expansion model with global rotation. However,
observations show that rotational effect can't be main cause of
the cosmological redshift. We consider metric for five-dimensional
space-time, which produce model of flat space with movement along
of additional coordinate. Kaluza- Klein theory proposes internal
spaces with Plank size to form an extra dimension. Besides this
approach, non-compactified interpretation of the fifth dimension
is treated \cite{4}. Machian interpretation of the Kaluza-Klein
theory is in \cite{5}. In framework of the brane world model the
possibility that 3-brane worldsheet moves in higher dimensional
space is discussed in \cite{6}. Cosmology of the Randall-Sundrum
(RS) model, considered as alternative of the compactification,
with radion stabilization is analyzed in \cite{7}. Charged Black
Holes \cite{8} and dynamical effects at the brane \cite{9} are
studied in generalized RS model. The fifth dimension was
introduced for unification gravitational and electromagnetic
fields \cite{10}, \cite{4}. Geodesic deviation in multidimensional
theories in presence of the Lorenz force is considered in
\cite{11}. On physical interpretation of the fifth dimension
\cite{4}, \cite{12}, \cite{13} the rest mass $m$ changed with time
is assumed to be $m\sim y$, where $y$ is fifth coordinate.
Multidimensional theories yield appearance in four-dimensional
space-time extra forces \cite{14}, \cite{15}. Analysis of the bulk
geodesic motion of the particle under influence of the extra
non-gravitational force in brane world \cite{16} admits the energy
generation, caused by change of the proper mass.

\section{5D field equations and its solution}

  Via a choice of units the speed of light $c$ and gravitational constant
$G$ are rendered to $c=1$ and $8\pi G=1$. The field equations are
taken to be 5D version \cite{17} of the usual 4D Einstein field
equations, namely
 \begin{equation}\label{f1}
 \mathbf{G}^{\mathrm{ij}}=\mathbf{T}^{\mathrm{ij}},
 \end{equation}
where $\mathbf{G}^{\mathrm{ij}}$ is 5D analogy of the Einstein
tensor, and $\mathbf{T}^{\mathrm{ij}}$ is 5D extension of 4D
energy-momentum tensor taken in general form \cite{18}. With
coordinates $x^{\mathrm{0}}=t$ (time),
$x^{\mathrm{1,2,3}}=\eta^{\mathrm{1,2,3}}$ (spacelike coordinates)
and $x^{\mathrm{4}}=y$ (extra coordinate) the line element is
given by
 \begin{equation}\label{f2}
 ds^2  = dt^2+2B(t)dtdy-B(t)^{2q}d\eta^{\mathrm{i}2} +
 2AB(t)^{1+q}d\eta^{\mathrm{1}}dy + (1-h^2)B(t)^2dy^2 ,
 \end{equation}
where $B(t)$ is function of $t$, $h$, $q$ and $A$ are constant
parameters, for definiteness $B(t)>0$. In order to avoid the
existence of closed timelike curves is required $h>1$. The case
$q\neq 0$ correspond to the space with changed scale factor, with
$q=0$ 3D space is stationary.

 With denotation $F(t)\equiv(\partial B/\partial t)/B$ and
$Q(t)\equiv (\partial^2 B/\partial t^2)/B$ for $q$ being equal $1$
or $0$ the non-vanishing field equations (\ref{f1}) are following:
\begin{eqnarray}\label{f3}
 \mu_g =[(1+47q)A^2 +(1+23q)A^4 -(1+47q)h^2 A^2 +24q(1 - 2h^2 
\nonumber \\
 +h^4 )]F^2 /[4(A^2 -h^2)^2], \\
\nonumber \\
 \mathbf{S}_{\mathrm{1}}=A[A^2 (1+ 23q)+24q(1-h^2)F^2]/[4(A^2 
 -h^2)^2 B^q ],   \\
\nonumber \\
 \mathbf{S}_{\mathrm{4}}=-[A^2(1+23q)+24q(1-h^2)]F^2/
 [4(A^2-h^2 )^2 B],   \\ 
\nonumber \\
 \mathbf{\sigma}_{\mathrm{11}}=([A^2 h^2 (1+11q)+12q(A^2+h^2-h^4)]F^2 
\nonumber \\
 +4(1+2q)[-A^2+h^2 +A^2 h^2-h^4]Q)/[4(A^2 -h^2)^2 B^{2}],       \\
\nonumber \\
 \mathbf{\sigma}_{\mathrm{22}}=\mathbf{\sigma}_{\mathrm{33}}= ([A^2
 (1-13q)+12q(-1+h^2)]F^2- 
\nonumber \\
4(1+2q)[1+A^2-h^2]Q)/[4(A^2 -h^2)B^{2q}], \\
\nonumber \\
\mathbf{\sigma}_{\mathrm{14}}=A([(-1+13q)A^2 +2(1-7q)h^2+24q]F^2+
\nonumber \\
2(1+5q)[A^2-h^2]Q)/[4(A^2-h^2)B^{1+q}],                        \\
\nonumber \\
 \label{f9}
 \mathbf{\sigma}_{\mathrm{44}}=([A^2(1+11q)+24-12h^2]F^2+12q[A^2
 -h^2]Q)/[4(A^2-h^2)^2B^2],
 \end{eqnarray}
where $\mu_g$  is mean energy density in the universe,
$\mathbf{S}_{\mathrm{i}}$  are components of vector of the energy
flow, $\mathbf{\sigma}_{\mathrm{ij}}$  is the 4D analogy of the 3D
strain tensor. Energy conservation law
$\partial\mathbf{T}^{\mathrm{ij}}/\partial x^{\mathrm{i}}=0$
yields relation
 \begin{equation}\label{f10}
 \frac{\partial \mu_g}{\partial t}+\frac{ \partial \mathbf{S}_{\mathrm{1}}}
 {\partial t}+ \frac{ \partial \mathbf{S}_{\mathrm{4}}}{\partial t}=0.
 \end{equation}

\section{Geodesic motion}

   Denoting five-velocities as $u^{\mathrm{i}}=dx^{\mathrm{i}}/ds$ let
us find a solution of the geodesic line equations for particle
with non-zero rest mass in form
 \begin{equation}\label{f11}
 \frac{d}{ds}(\mathbf{g}_{\mathrm{ij}}u^{\mathrm{j}})-\frac{1}{2}\frac{
 \partial \mathbf{g}_{\mathrm{mj}}}{\partial x_{\mathrm{i}}}u^{\mathrm{m}}
 u^{\mathrm{j}}=0 ,
 \end{equation}
where $\mathbf{g}_{\mathrm{ij}}$ is metrical tensor. Spatial
components are satisfied with comoving coordinates of the
conventional type $u^{\mathrm{1}} =u^{\mathrm{2}} =u^{\mathrm{3}} =0$.
Then zero, first and fourth components of the equations
(\ref{f11}) yield
 \begin{eqnarray}\label{f12}
 \frac{d}{ds}(u^{\mathrm{0}}+Bu^{\mathrm{4}})-\frac{\partial B}{\partial t}
 u^{\mathrm{0}}u^{\mathrm{4}}-(1-h^2)B\frac{\partial B}{\partial t}
 u^{\mathrm{4}}u^{\mathrm{4}}=0, \nonumber \\
 \frac{d}{ds}(AB^{1+q}u^{\mathrm{4}}) = 0 ,   \nonumber     \\
 \frac{d}{ds}[Bu^{\mathrm{0}} + (1-h^2)B^2 u^{\mathrm{4}}] = 0.
 \end{eqnarray}
 A solution of these equations has to be compatible with the
condition set by metric (\ref{f2}), which is
 \begin{equation}\label{f13}
 (u^{\mathrm{0}} + Bu^{\mathrm{4}})^2-h^2B^2 (u^{\mathrm{4}})^2=1.
 \end{equation}
In cases $q=0$ and $q\neq 0$, $A=0$ equations (\ref{f12})
not constraining the signs of the velocities turn out to
 \begin{eqnarray}\label{f14}
 u^{\mathrm{0}}=\pm (h^2-1)^{0.5}/h,\nonumber  \ \
 u^{\mathrm{4}}=\pm 1/[Bh(h^2-1)^{0.5}].
 \end{eqnarray}
The ratio of these equations gives
 \begin{equation}\label{f15}
 dy/dt =1/[(h^2 -1)B] .
 \end{equation}

\section{The Pioneer effect}

  In terms of the action of the system along of the world line, Lagrangian
is written as
 \begin{equation}\label{f16}
 L= -m[1-B^{2q}\dot{\eta}^{\mathrm{i}2} +2B\dot{y}
 +2AB^{1+q}\dot{\eta}^{\mathrm{1}}\dot{y}+(1-h^2)B^2\dot{y}^2]^{0.5} ,
 \end{equation}
where $m$ is the rest mass of the particle, and overdot denotes
derivative with respect to $t$. Proposed model suggests
explanation of the Pioneer effect \cite{19}, \cite{20}. In
comoving coordinate system with $\dot{y}=0$ for located at the
distance $r$ particle, moved slowly in weak gravitation field,
Lagrangian (\ref{f16}) turn out
 \begin{equation}\label{f17}
 L =-m[1-\frac{1}{2}B^{2q}\dot{\eta}^{\mathrm{i}2}+
 B\delta \dot{y}+AB^{1+q}\dot{\eta}^{\mathrm{1}}\delta\dot{y}
 + \frac{1}{2}(1-h^2)B^2 \delta \dot{y}^2+\varphi],
 \end{equation}
where $\varphi$ is gravity potential, $\delta \dot{y}$  is
difference of the velocity along of the fifth coordinate on the
light cone and in locally scaled coordinates, which is assumed to
be small. Taking into account relation (\ref{f15}) for geodesic
line, for small variation $B$, change of $\dot{y}$  is given by
 \begin{equation}\label{f18}
 \delta \dot{y}=F\delta t/[(h^2-1)B] ,
 \end{equation}
where $\delta t$ is time of the passing of the light of the
distance from particle. The scale factor in case $q=1$ during time
interval $\delta t$ is assumed to be $B\approx 1$. Then with
$|\delta \dot{y}|<<|\dot{\eta}^{\mathrm{i}}|$ equation (\ref{f17})
yields
 \begin{equation}\label{f19}
  L=-m[1- \frac{1}{2}\dot{\eta}^{\mathrm{i}2}+ F\delta t/(h^2-1)+\varphi].
 \end{equation}
The last two terms in (\ref{f19}) are considered as potential
energy of the system
 \begin{equation}\label{f20}
 U = m[Fr/(h^2-1)+\varphi],
 \end{equation}
where $r$ corresponds to the distance from Sun to the spacecraft.
Besides the gravity, one determine the additional force, which
yields for $F>0$ extra radial acceleration towards the Sun
 \begin{equation}\label{f21}
 a_p = F/(h^2-1).
 \end{equation}
This acceleration amounted $7.5\cdot 10^{-8} \ \mathrm{sm/s}^2$
\cite{20} takes place without dependence on distance, direction of
the radial movement and its and angular velocities that consists
with the Pioneer effect.

\section{Cosmological models}

  For expanding 3D space the Hubble parameter of the cosmological
redshift is assumed to be
 \begin{equation}\label{f22}
 H = F.
 \end{equation}
Then taking into account constrains for $H$ \cite{21} equation
(\ref{f21}) leads to $h\approx\sqrt{2}$ . With $A=0$ equations
(\ref{f3})-(\ref{f9}) yield isotropic model of the universe with
energy density $\mu_g\approx\frac{3}{2}H^2$ . The same field
equations admit anisotropic stationary model with $A\neq 0$,
however, in this case the cause of the cosmological redshift is
not clear.

\section{Conclusion}

Presented five-dimensional cosmological model allows explanation
of additional force produced Pioneer effect as a result of the
movement of the 3D space along extra coordinate. This approach
admits existence both isotropic expanded and anisotropic
stationary universes, however, the first appears to be more
probable.

\subsection*{Acknowledgements}

This work was supported by MSO.

\end{document}